\documentclass[10pt,preprint]{aastex}
\shortauthors{Gabel et al.}
\shorttitle{Averaged STIS and {\it FUSE} Spectra of NGC 3783}
\received{2002 June 12}
\begin{document}
\title{The Ionized Gas and Nuclear Environment in NGC 3783\\
II. Averaged {\it HST}/STIS and {\it FUSE} Spectra\altaffilmark{1}}
\author{Jack R. Gabel\altaffilmark{2}, D. Michael Crenshaw\altaffilmark{3}, 
Steven B. Kraemer\altaffilmark{4}, W. N. Brandt\altaffilmark{5}, Ian M. George\altaffilmark{6,7},
Frederick W. Hamann\altaffilmark{8}, Mary Elizabeth Kaiser\altaffilmark{9}, Shai Kaspi\altaffilmark{5,10}, 
Gerard A. Kriss\altaffilmark{9,11}, Smita Mathur\altaffilmark{12},
Richard F. Mushotzky\altaffilmark{6}, Kirpal Nandra\altaffilmark{6,13}, Hagai Netzer\altaffilmark{10}, 
Bradley M. Peterson\altaffilmark{12}, Joseph C. Shields\altaffilmark{14},
T. J. Turner\altaffilmark{6,7}, \& Wei Zheng\altaffilmark{9}}
\altaffiltext{1}{Based on observations made with the NASA/ESA {\it Hubble 
Space Telescope}.  STScI is operated by the Association of Universities 
for Research in Astronomy, Inc. under NASA contract NAS~5-26555.}
\altaffiltext{2}{The Catholic University of America/IACS, NASA/Goddard Space Flight Center, Laboratory for Astronomy and Solar Physics, Code 681, Greenbelt, MD 20771.}
\altaffiltext{3}{Department of Physics and Astronomy, Georgia State University,
Atlanta, GA 30303.}
\altaffiltext{4}{The Catholic University of America, NASA/Goddard Space Flight Center, Laboratory for Astronomy and Solar Physics, Code 681, Greenbelt, MD 20771.}
\altaffiltext{5}{Department of Astronomy and Astrophysics, 525 Davey Laboratory, The 
Pennsylvania State University, University Park, PA 16802.}
\altaffiltext{6}{Laboratory for High Energy Astrophysics, NASA/Goddard Space Flight Center,
Code 662, Greenbelt, MD 20771.}
\altaffiltext{7}{Joint Center for Astrophysics, Physics Department, University of Maryland,
Baltimore County, 1000 Hilltop Circle, Baltimore, MD 21250.}
\altaffiltext{8}{Department of Astronomy, University of Florida, 211 Bryant Space Science Center,
Gainesville, FL, 32611-2055.}
\altaffiltext{9}{Center for Astrophysical Sciences, Department of Physics and Astronomy,
The Johns Hopkins University, Baltimore, MD 21218-2686.}
\altaffiltext{10}{School of Physics and Astronomy, Raymond and Beverly Sackler Faculty of
Exact Sciences, Tel-Aviv University, Tel-Aviv 69978, Israel.}
\altaffiltext{11}{Space Telescope Science Institute, 3700 San Martin Drive, Baltimore, MD 21218.}
\altaffiltext{12}{Department of Astronomy, Ohio State University, 140 West 18th Avenue,
Columbus, OH 43210-1173.}
\altaffiltext{13}{Universities Space Research Association, 7501 Forbes Boulevard, Suite 206, 
Seabrook, MD 20706-2253.}
\altaffiltext{14}{Department of Physics and Astronomy, Clippinger Research Labs 251B, Ohio
University, Athens,  OH 45701-2979.}

\begin{abstract}
     We present observations of the intrinsic absorption in the Seyfert 1 galaxy
NGC 3783 obtained with the Space Telescope Imaging Spectrograph (STIS) on the 
{\it Hubble Space Telescope} ({\it HST\,}) and the {\it Far Ultraviolet Spectroscopic Explorer}
({\it FUSE\,}).  We have combined 18 STIS and 5 {\it FUSE} observations to obtain
a high signal-to-noise averaged spectrum spanning 905--1730~\AA .  The averaged spectrum reveals
absorption in \ion{O}{6}, \ion{N}{5}, \ion{C}{4}, \ion{N}{3}, \ion{C}{3} and the Lyman lines 
up to Ly$\epsilon$ in the three blueshifted kinematic components previously detected in the STIS 
spectrum (at radial velocities of $-$1320, $-$724, and $-$548~km~s$^{-1}$).  The highest 
velocity component exhibits absorption in \ion{Si}{4}.
We also detect metastable \ion{C}{3} in this component, indicating a high density in this absorber.  
No lower ionization lines, i.e., \ion{C}{2} and \ion{Si}{2}, are detected.  A weak, 
fourth absorption component is tentatively detected in the high ionization lines and Ly$\alpha$ and 
Ly$\beta$ at a radial velocity of $-$1027~km~s$^{-1}$.  The Lyman lines reveal a complex absorption geometry.  
The strength of the higher order lines indicates Ly$\alpha$ and Ly$\beta$ are saturated over much of the resolved 
profiles in the three strongest absorption components and, therefore, their observed profiles are 
determined by the covering factor.  We separate the individual covering factors of the continuum 
and emission-line sources as a function of velocity in each kinematic component using the Ly$\alpha$ 
and Ly$\beta$ lines.  The covering factor of the BLR is found to vary dramatically between the cores 
of the individual kinematic components, ranging from 0 to 0.84.  Additionally, we find that the 
continuum covering factor varies with velocity within the individual kinematic components, decreasing 
smoothly in the wings of the absorption by at least 60\%. 
Comparison of the effective covering factors derived from the \ion{H}{1} results with those determined 
directly from the doublets reveals the covering factor of \ion{Si}{4} is less than half that of 
\ion{H}{1} and \ion{N}{5} in the high velocity component.  Additionally, the FWHM of \ion{N}{3} 
and \ion{Si}{4} are narrower than the higher ionization lines in this component.  These results 
indicate there is substructure within this absorber.  We also find evidence for structure in the 
column density profiles of the high ionization lines in this component.  We derive a lower limit on 
the total column ($N_H \geq$~10$^{19}$~cm$^{-2}$) and ionization parameter ($U \geq$~0.005) in the 
low ionization subcomponent of this absorber.  
The metastable-to-total \ion{C}{3} column density ratio implies $n_e \approx$~10$^9$~cm$^{-3}$
and an upper limit on the distance of the absorber from the ionizing continuum of 
$R \leq$~8~$\times$~10$^{17}$~cm. 
The decreasing covering factor found in the wings of the absorption and the extreme compactness of
the C III$^*$ absorber are suggestive of a clumpy absorption gas with low volume filling factor.
\end{abstract}
\keywords{galaxies: individual (NGC 3783) --- galaxies: active --- galaxies: Seyfert --- ultraviolet: galaxies}


\section{Introduction}      
     Observations with the {\it Hubble Space Telescope} ({\it HST\,}) have 
revealed that intrinsic UV absorption is a common phenomenon in Seyfert 1
galaxies, appearing in over half of the objects with available spectra
\citep{cren99a}.  The UV absorption resonance lines are typically blueshifted in the
rest frames of the host galaxies, indicating radial outflow.  The absorption
is often highly variable and, in some cases, only partially covers the
continuum and emission-line regions of the active nuclei, which implies
the absorbers are intrinsic to the AGN environments.  The observed variations
may be due to changes in ionization in the absorbers \citep{krol97,shie97,cren00} or changes in the total
absorbing column, i.e., as a result of motion into and out of our
line-of-sight \citep{cren99b}.

     NGC~3783 is a bright Seyfert 1 galaxy that exhibits strong UV absorption 
features and X-ray ``warm absorption".  Several observations with {\it HST} 
over the past decade have revealed dramatic variability in the UV absorption.
NGC~3783 also has a highly variable UV continuum source.  An {\it International Ultraviolet 
Explorer} ({\it IUE\,}) monitoring campaign revealed a 
factor of $\sim$2 flux variations over timescales of 20--40 days \citep{reic94,onke02}.

   Intrinsic absorption in Ly$\alpha$ and \ion{C}{4}~$\lambda\lambda$1548,1551 
was first detected in NGC~3783 with Faint Object Spectrograph (FOS) observations 
by \citet{reic94}.  Three subsequent spectra obtained over a period 
of about two years with the Goddard High Resolution Spectrograph (GHRS) 
revealed highly variable \ion{C}{4} absorption \citep{mara96,cren99a}.
There was no detectable \ion{C}{4} absorption in the 1993 February GHRS spectrum,
however, by 1995 April, two kinematic components appeared, at radial velocities of 
$-$1365~km~s$^{-1}$ (referred to as component 1) and $-$548~km~s$^{-1}$ (component 2) 
relative to the systemic redshift \citep[we adopt $z =$~0.009760~$\pm$~0.000093 throughout
this paper;][]{deva91}.
Additionally, a GHRS spectrum of the \ion{N}{5}~$\lambda\lambda$1239,1243 spectral region revealed 
absorption coincident
in velocity with component 2 just 16 days after the 1993 February observation
of \ion{C}{4} that showed no absorption, suggesting rapid variability \citep{lusa94}.
A Space Telescope Imaging Spectrograph (STIS) medium 
resolution echelle spectrum of NGC 3783 was obtained on 2000 February 27 revealing a 
third kinematic component in \ion{C}{4} ($v_r=-$724 km s$^{-1}$, component 3), in 
addition to the components seen in the final GHRS spectrum \citep{krae01a}.  
Ly$\alpha$ and \ion{N}{5}~$\lambda\lambda$1239,1242 also appeared in these 
three kinematic components in the STIS spectrum.  \ion{Si}{4}~$\lambda\lambda$1394,1403 
absorption was only found in component 1 and no lines from lower ionization species, 
i.e., \ion{C}{2}, \ion{Si}{2}, Mg II, were detected.  Using the \ion{N}{5} doublet 
lines, \citet{krae01a} found that all three absorption systems have a non-unity 
effective covering factor.  No correlation was found between the strength of the UV 
absorption features and continuum flux in the GHRS and STIS spectra by \citet{krae01a}. 
They concluded the observed absorption variations were due largely to a change in
total column.  However, it was not possible to constrain tightly the variation timescales
due to the sampling of these observations.  Additionally, these observations did not sample the rapid 
changes in the continuum flux that were observed in the study by \citet{reic94}, thus, it 
remains unclear what affect variable ionization has on the absorption.

     A large ($>$~10$^{22}$ cm$^{-2}$) and variable column of ionized gas was measured 
in the X-ray spectrum of NGC~3783 with the {\it Advanced Satellite for Cosmology and
Astrophysics} ({\it ASCA\,}) \citep[e.g.,][]{geor98}.  Subsequent observations with the High 
Energy Transmission Grating Spectrometer (HETGS) aboard the {\it Chandra X-ray Observatory}
({\it CXO\,}) showed numerous absorption lines with a mean radial velocity of 
$\sim-$610~$\pm$~130~km~s$^{-1}$, consistent with components 2 and 3 identified 
in the UV \citep{kasp00,kasp01}.  The coincidence in velocities suggests a link between 
the UV and X-ray absorbers.  Photoionization modeling of the STIS spectrum revealed that, 
although the UV absorbers can produce some of the observed X-ray columns, the ionization 
is too low and the total column too small to account for all of the features in the X-ray 
\citep{krae01a}.  Hence, the exact relationship of the UV and X-ray absorption remained 
uncertain.  Additionally, \citet{krae01a} found that two zones are required
to explain the UV absorption columns measured in component~1.  The strength of 
the \ion{Si}{4} absorption is inconsistent with the large \ion{N}{5}/\ion{C}{4} column 
density ratio measured in the STIS spectrum.  They proposed a model where a 
relatively high density absorber is co-located with more tenuous gas that is more
highly ionized.

     Various studies of intrinsic absorption in AGNs have demonstrated the 
importance of determining the absorption covering factor in the line-of-sight to the nucleus.  
In addition to affecting column density measurements, the covering factor constrains the 
absorption and emission geometry, providing tests for physical models.
Non-unity effective covering factors, $C_f$, can arise from scattering of light into our 
line-of-sight by an extended scatterer \citep[e.g.,][]{cohe95,good95,krae01b} or partial 
occultation of the emission sources \citep{wamp93,barl97a,hama97}.  In the most general case, the 
different covering factors of the continuum and emission-line region need to be 
determined \citep{barl97a,gang99}, and $C_f$ can vary 
across the profile of an individual absorption component \citep[e.g.,][]{barl97b,arav02}. 
Additionally, instrumental scattering can affect the derived $C_f$ and column
densities \citep{cren98}.  High-resolution spectra with high signal-to-noise 
(S/N) are required to separate these effects.
 
     We have undertaken an intensive multiwavelength monitoring campaign 
to probe the intrinsic absorption in NGC 3783 using observations in the 
UV, far-UV, and X-ray.  In this paper, we present an analysis of the averaged 
spectrum in the UV and far-UV from STIS and the {\it Far Ultraviolet Spectroscopic
Explorer} ({\it FUSE\,}).  We have co-added all observations to produce a high
S/N spectrum that samples numerous lines from a range of ionization states.  Analysis of 
the mean X-ray spectrum is presented in \citet{kasp02}, hereafter Paper~I.
Details of the variable nature of the absorption will be presented in 
future papers by Gabel et al.\ (in preparation) and George et al.\ (in preparation).  
This paper is organized as follows: in \S 2, we present details of the 
STIS and {\it FUSE} observations and data reduction; in \S 3, measurements
of the covering factors and column densities for the UV absorbers are
given; in \S 4 we derive constraints on the physical conditions and geometry
of the absorbers; we summarize our results in \S 5.

\section{Observations, Data Reduction, and the UV -- Far-UV Spectrum}  
     The nucleus of NGC~3783 was observed with {\it HST}/STIS at 18 epochs between
2000~February~27 to 2002~January~6, and with {\it FUSE} at 5 epochs between 2001~February~28 to 
2001~June~30.  Each STIS medium-resolution echelle spectrum was obtained in two orbits through the 0$\farcs$2~$\times$~0$\farcs$2 aperture centered on the nucleus using the E140M grating, which 
covers 1150-1730~\AA .  Exposure times and observation dates are given in Table 1.  All spectra 
were reduced using the IDL software developed at NASA Goddard Space Flight Center 
for the Instrument Definition Team.  We incorporated a procedure to remove the background 
light from each order using a scattered light model devised by \citet{lind99}.  Each echelle 
spectrum was generated by averaging together the individual orders in regions of overlap. 
The extracted STIS spectra are sampled in $\sim$5~km~s$^{-1}$ bins.
We measured residual fluxes in the cores of saturated Galactic lines in the STIS spectra (Ly$\alpha$,
\ion{O}{1}~$\lambda$1302, \ion{C}{2}~$\lambda$1335, \ion{Si}{2}~$\lambda$1527) to
test the accuracy of the removal of scattered light.
We find in all cases mean flux levels measured over the cores are consistent with zero within the 
noise levels.  Typical values for the mean fluxes in the cores are $\pm$ 0--3 \% of the local
continuum flux, compared to uncertainties due to noise (i.e., the standard deviations
measured over these intervals) of $\sim$5--12 \% of the local continuum flux.

     The {\it FUSE} spectra, which cover 905-1187~\AA, were obtained in time-tag mode with the 30$\arcsec$~$\times$~30$\arcsec$ low-resolution aperture (LWRS).  Each observation consists of 
multiple exposures obtained in consecutive orbits on four detector segments, with two spectra (from the SiC 
and LiF channels) on each segment.  Details of the observations are given in Table~2.  Using IDL 
software obtained from the {\it FUSE} data analysis web page 
(http://fuse.pha.jhu.edu/analysis/ analysis.html), we examined each raw two-dimensional image for 
instrumental effects that may contaminate the data.  Numerous ``event bursts" were found 
\citep[see discussion in][]{oege00} and an instrumental artifact affects the long wavelength region 
of all LiF~1B spectra.  This artifact, known as ``the worm", is due to shadowing by one of the grid wires 
above the detector \citep{oege00}.  The raw data were screened to remove these effects and 
processed using the standard {\it FUSE} calibration pipeline, CALFUSE v1.8.7.  
     
     CALFUSE extracts eight one-dimensional spectra for each exposure, 
corresponding to the two channels and four detector segments.  
Since the extracted {\it FUSE} spectra are oversampled, we rebinned to 
8~km~s$^{-1}$~bin$^{-1}$ to achieve higher S/N while preserving the full instrumental 
resolution (one resolution element corresponds to two data bins), which
is nominally $\sim$20~km~s$^{-1}$ \citep{oege00}.
For each channel/segment, we compared the wavelength scales for all exposures using 
cross-correlation and corrected for any offset.  All exposures 
were then co-added, weighted by exposure time.  To achieve the highest possible S/N, 
the resulting spectra for the eight channel/segments were also co-added.
To correct for the non-linear shifts in wavelength scale in the
detector segments, we cross-correlated and aligned the spectra over small bandpasses. 
The spectra were then co-added by weighting each channel/segment with its effective
area, using the effective area versus wavelength functions given in \citet{blai00}.
The absolute wavelength scale was determined from the intrinsic wavelengths of the Galactic lines.
We measured flux levels in the cores of saturated Galactic H$_2$ lines, and find they
are consistent with zero within the noise of the data.

\subsection{The Averaged UV -- Far-UV Spectrum}

     The individual STIS and {\it FUSE} observations were co-added to 
produce an averaged spectrum for each instrument.  We have tested the individual observations
extensively and find that our results in this paper are not strongly sensitive to variability in the
spectra.  Specifically, we performed all measurements described in the subsequent analysis on each
STIS and {\it FUSE} observation, as well as on various coadded combinations of the individual observations.
We find that the measurements of the covering factors in \S 3.1 and \S 3.2 below are not affected
by variability, within the uncertainties of those measurements.  There are variations in column densities for some of the ions, and these will be given a full treatment in an upcoming paper (Gabel et al.\ [in preparation]).
Thus, the column density measurements presented in \S 3.2 are average values.

     In Figure 1, the spectra for the \ion{O}{6}+Ly$\beta$, Ly$\alpha$+\ion{N}{5}, \ion{C}{4}, and \ion{Si}{4} resonance lines are plotted.  Wavelengths corresponding to four previously identified intrinsic absorption components are denoted with tick marks for each line.  Components 1--3 are
those identified in the initial STIS spectrum by \citet{krae01a}.
Component 4, at v$_r=-$1027~km~s$^{-1}$ in the rest frame of the host galaxy, was first identified 
in \ion{O}{6} and Ly$\beta$ in an earlier {\it FUSE} spectrum \citep{kais02,gabe02}.  
We note that absorption in this component is weak or absent in all other lines
and its detection is very sensitive to the fit to the intrinsic emission.  
Additionally, Galactic \ion{O}{6}~$\lambda$1032 absorption may contribute to the observed Ly$\beta$ feature
(this will be addressed further in a study of the Galactic absorption by Mathur et al.\ [in preparation]).
Given these uncertainties, we consider component~4 to be a tentative detection of intrinsic absorption.
Interstellar absorption lines and detector artifacts are also identified on the top of each spectrum in 
Figure 1.  The local interstellar absorption includes three kinematic components, at 0, +60, and 
+240 km~s$^{-1}$ with respect to the Galactic rest frame \citep{lusa94}.  

     We modeled the intrinsic continuum and emission line fluxes at each absorption feature
in the following manner ("intrinsic'' flux here refers to the AGN emission before 
passing through the absorbers).
First, we modeled the intrinsic continuum {\it plus} emission-line flux
by fitting a cubic spline to unabsorbed regions of the spectrum on either side of the line. 
Next, to determine the relative contribution of continuum and
line emission at each absorption feature, we estimated the {\it continuum} flux levels 
at each absorption feature by measuring the flux in regions of the spectrum that
are not affected by line emission and linearly interpolating between these regions.  
Due to heavy contamination of the Ly$\alpha$ and \ion{O}{6} spectral regions by Galactic absorption,
we used the \ion{C}{4} emission-line profile as a template in fitting the intrinsic emission
for these lines.  Specifically, we subtracted our fit to the continuum flux from our fit to the total 
\ion{C}{4} emission spectrum to derive its emission-line profile.  We then scaled this profile 
in flux, preserving its velocity, and added it to the continuum flux for Ly$\alpha$ and 
\ion{O}{6} to derive the fit to these lines.  We find that the \ion{C}{4} line template 
does not fit the \ion{O}{6} emission profile fully; an additional, narrow emission 
component is required.  We modeled this with a FWHM~$=$~350~km~s$^{-1}$ Gaussian profile 
centered at the systemic velocity of NGC~3783.  Since this narrow line region component
does not appear in the emission-line profiles in the STIS spectrum, it
likely arises in an extended region that is not fully covered by the 0$\farcs$2~$\times$~0$\farcs$2 
STIS aperture.  The broad line region (BLR) emission from Ly$\beta$ and higher order Lyman lines is 
negligible in the {\it FUSE} spectrum.  We have generated comparison models using the 
photoionization code CLOUDY \citep{ferl98} for typical conditions in BLR gas and find this 
is consistent with the predicted spectrum.  Multiple absorptions of the higher order Lyman 
line photons followed by cascade produce an \ion{H}{1} spectrum dominated by Ly$\alpha$.
Thus, the line emission that is absorbed by Ly$\beta$ is high velocity \ion{O}{6} BLR emission.  
Our fits to the total (dashed line) and continuum (dot-dashed line) fluxes are plotted on the spectra in 
Figure~1.  We have modeled the Galactic H$_2$ and \ion{H}{1} absorption, using the parameters derived
in \citet{semb01} for the high velocity cloud component.  The fits to \ion{O}{6}, Ly$\beta$ and 
Ly$\alpha$ shown in Figure~1 include absorption by this model.  

     In Figure~2, we show the spectra of Ly$\beta$ and \ion{O}{7} illustrating the kinematic
link betweeen the UV and X-ray absorption.  The \ion{O}{7} profile is a combination of the four strongest
lines from that ion in the mean {\it Chandra} spectrum (see Figure~10 in Paper~I).  The velocities of the four 
intrinsic UV absorption components are marked with tick marks above the spectra.  
Structure in the X-ray absorption profiles is revealed in the high S/N mean {\it Chandra} spectrum.  In Figure~2, at least two kinematic components are apparent in the \ion{O}{7} profile, coinciding with UV components~1 and 2+3 (see discussion in \S 3.5 of Paper~I).  We cannot rule out the presence of \ion{O}{7} absorption in component 4 as well.
  
\section{Measurements}   
     We normalized all absorption features by dividing the averaged spectrum by our fits to the 
intrinsic emission (see \S 2.1).  The normalized profiles are plotted as a function of radial velocity in Figure~3.  The centroid velocities of the four intrinsic absorption components 
are marked with dashed vertical lines on each spectrum.  These were derived from the \ion{N}{5}
absorption for components~1--3 and Ly$\beta$ for component~4.   
We have marked interstellar lines and known detector artifacts (labeled with "art") with short tick marks in Figure~3. 
Qualitatively, the absorption in the merged STIS data is similar to 
the initial spectrum presented in \citet{krae01a}.  Ly$\alpha$, \ion{N}{5}, and \ion{C}{4}
are strong in components 1--3, while \ion{Si}{4} is only detectable in component~1.  
The \ion{C}{4}~$\lambda$1551 line in component~3 is blended with \ion{C}{4}~$\lambda$1548 in 
component~1.  The blue wing of \ion{N}{5}~$\lambda$1242 component~1 is blended
with the red wing of \ion{N}{5}~$\lambda$1238 component~2.

     In the {\it FUSE} spectrum, the \ion{O}{6}~$\lambda$1032 line is heavily contaminated 
with Galactic features, but it appears in all components in \ion{O}{6}~$\lambda$1038.
The deep feature in the red wing of component~1 in \ion{O}{6}~$\lambda$1038 
is a detector artifact.  We detect absorption in the Lyman lines up to Ly$\epsilon$ in components 1--3.  
Higher order Lyman lines may be present, but suffer from severe blending with
Galactic absorption.  Additionally, strong Galactic absorption contaminates component~1 in Ly$\gamma$
and component~2 in Ly$\epsilon$.  In the top window of Figure~3a, Ly$\beta$ is 
seen to be deeper than Ly$\alpha$ in components 1, 3, and 4, despite having an optical depth that is six 
times smaller.  Additionally, the depth of Ly$\epsilon$ in components 1 and 3 indicates Ly$\beta$ 
is saturated in these components since $\tau_\beta$/$\tau_\epsilon=$~11, but the residual flux
levels in the cores of the Ly$\beta$ features are non-zero.  These are signatures of a complex 
absorption geometry in NGC~3783, which is addressed further in \S 3.1.
Both \ion{N}{3}~$\lambda$989 and \ion{N}{3}$^* \lambda$991, which arises from the
fine-structure level, are strong in component~1.
The \ion{N}{3}~$\lambda$989 component~1 profile shown in Figure~3c exhibits a sharp
edge in the blue wing and spiked core blueward of the centroid.  
These features are not seen in the profiles of any other lines in this component, most
notably the \ion{N}{3}$^* \lambda$991 line.
Furthermore, there is a strong Galactic H$_2$ line at this wavelength contaminating the \ion{N}{3} absorption.
Thus, this spike is likely due to noise in the deep trough of the H$_2$ + \ion{N}{3} absorption.
Weak absorption is also detected in components~2 and 3 in the \ion{N}{3}$^* \lambda$991 line.
\ion{C}{3}~$\lambda$977, also predicted to be strong in component 1 \citep{krae01a}, is 
heavily blended with Galactic H$_2$ and cannot be measured.  However, the 
\ion{C}{3}$^* \lambda\lambda$1175--76 multiplet, which arises from metastable levels 6.5 eV above 
ground, is detected in the STIS spectrum, indicating high density in this component 
\citep{brom85}.  To our knowledge, this is the first detection of \ion{C}{3}$^* \lambda\lambda$1175--76
absorption in a Seyfert galaxy other than NGC~4151.  Weak \ion{C}{3}~$\lambda$977 
absorption also appears in components 2 and 3.  There is a weak feature corresponding to the \ion{P}{5}~$\lambda$1118 line at the velocity of component~1.  However, we cannot confirm the presence of this component in the
other line of the doublet, \ion{P}{5}~$\lambda$1128, due to contamination with Galactic absorption.
The \ion{S}{6}~$\lambda\lambda$933,944 resonance doublet 
is contaminated with Galactic interstellar features for all components and cannot be measured.  Lines of lower ionization, such as \ion{C}{2} and \ion{Si}{2}, are not detectable in the spectrum.  

\subsection{Covering Factor Analysis of the Lyman Lines}
     The observed absorption depths are determined by a combination of the 
effective covering factor and the optical depths of the lines.  The absorption 
must be properly corrected for $C_f$ before an accurate column density can be 
measured.  In general, the different covering factors of the line and continuum
sources, $C_f^l$ and $C_f^c$, must be taken into account.  Following the method
of \citet{gang99}, we assume the same column of gas obscures the two emission sources, 
i.e., $\tau_l = \tau_c$. 
The observed residual flux in any absorption line is then given by
\begin{equation}
F = F_l  (C_f^l  e^{-\tau} + 1 - C_f^l) + F_c  (C_f^c  e^{-\tau} + 1 - C_f^c), 
\end{equation}
where $F_l$ and $F_c$ are the intrinsic line and continuum fluxes, respectively.  Dividing 
equation~1 by $F_l + F_c$ gives the expression for normalized flux,
\begin{equation}
I = R_l  (C_f^l  e^{-\tau} + 1 - C_f^l) + R_c  (C_f^c  e^{-\tau} + 1 - C_f^c), 
\end{equation}
where $R_l=F_l/(F_l+F_c)$ and $R_c=F_c/(F_l+F_c)$, which represent the relative 
contributions of line and continuum emission to the total intrinsic flux.

     The Lyman lines provide a unique study of the covering factor. 
They give a set of lines that (a) span a large range in optical depth and (b) have substantially 
different relative contributions from continuum and line emission, $R_c$ and $R_l$.  Since they arise 
from the same atomic energy level, their optical depths scale 
simply by their $f \lambda$ ratios, where $f$ is the oscillator strength and $\lambda$ the laboratory 
wavelength of the lines.  In Figure~3a, the strong absorption in the cores of components~1 and 3 in 
Ly$\epsilon$, and component~2 in Ly$\gamma$ (Ly$\epsilon$ is blended with Galactic \ion{C}{1} in this component) 
indicates both Ly$\alpha$ and Ly$\beta$ are saturated in these components and, therefore, {\it their core absorption depths are determined entirely by the covering factor}.  Since Ly$\alpha$ and Ly$\beta$ are 
saturated, $e^{-\tau} \approx$~0 for these two lines and their normalized flux given by equation~2 reduces 
to a simple function of $C_f^l$ and $C_f^c$.  We have combined these simplified equations for Ly$\alpha$ and 
Ly$\beta$ and solved for $C_f^l$ and $C_f^c$ in the cores of the absorption, thereby separating the individual covering factors of the line and continuum sources.  In these measurements, we averaged over 50~km~s$^{-1}$ intervals
centered on the centroid velocities of the components.  We note that this analysis assumes the $C_f^l$ of 
the \ion{O}{6} emission (which affects the Ly$\beta$ absorption profile) and the Ly$\alpha$ emission are 
identical.    

     In Table~3, we list the normalized residual flux, $I$, and the fractional contribution 
of continuum emission to the total intrinsic flux, $R_c$, measured in the core of each available 
Lyman line, along with our derived values for $C_f^l$ and $C_f^c$.  
The 1~$\sigma$ uncertainties quoted in Table~3 were computed from repeated measurements using different 
placements of intrinsic continuum and BLR flux levels, and include the uncertainty in the residual flux due to noise.  These results show the discrepancies in the apparent absorption strengths seen in 
Figure~3a are due to different covering factors of the line and continuum sources.  For example, in component~1 
where the Ly$\beta$ absorption is deeper than Ly$\alpha$, $R_c =$~0.57 and 0.15 for Ly$\beta$ and Ly$\alpha$, respectively, indicating $C_f^l < C_f^c$ in this component.

   A closer look at the Lyman profiles in Figure~3a reveals the absorption in the higher order 
lines is resolved in the individual kinematic components, and is deep over several resolution elements 
in the {\it FUSE} data.  Thus, we extended the above analysis to derive the line and continuum 
covering factors as a function of radial velocity.  An exact solution to $C_f^l$ and $C_f^c$ is obtained
at all radial velocities where Ly$\beta$ is saturated.  
Since the optical depths of the Lyman lines scale according to their $f \lambda$ ratios, this 
can be determined from the strengths of the higher order Lyman lines.
In general, $I \geq e^{-\tau}$, where the $>$ in the relation applies if the covering factor is less than
unity.  Thus, the velocities where $(I_{LyX})$$^{\tau_\beta/\tau_X}$ equals zero within uncertainties
gives a conservative estimate of the regions of saturated Ly$\beta$ (X represents $\epsilon$ for 
components~1 and 3 and $\gamma$ for component~2).
For velocities where Ly$\beta$ is not saturated, such as in component~4 and in the outer wings 
of the absorption in components 1--3, the solution gives only lower limits to $C_f^l$ and $C_f^c$.  
In Figure~4, the derived covering factor profiles are plotted.  Velocities with only lower limits on the 
covering factors are plotted with a dotted line.  Figure~4a shows that the continuum covering factor 
decreases smoothly in the blue wings of components 1 and 3, at velocities where the covering factor 
is well determined (solid line).  
Thus, the decreased absorption observed in these wings in Ly$\beta$ 
(Figure~3a) is a result of decreasing coverage of the continuum source, and the profile
shape traces out the unocculted flux.  
The other absorption wings may also exhibit decreased covering factor, however, only lower limits are available 
at these velocities.  
As discussed in \citet{gang99}, instrumental smearing can introduce an error into computations of
the covering factor profiles, particularly in the wings.  
To test this effect, we convolved the observed absorption profiles with the line spread function
of {\it FUSE}.  We find that the changes in the profiles are indetectable within the
noise of the data, and conclude that instrumental smearing has a negligible effect on our results.
The variable covering factor detected in this analysis indicates there is substructure in the individual 
absorbers and may provide a constraint on physical models of the absorption gas.  This is addressed
further in \S 4.

     One possible source of error associated with our calculations
that is not included in the uncertainties quoted in Table~3 is if the covering factor of
the \ion{O}{6} and Ly$\alpha$ BLRs are not the same.  The intensive {\it IUE} monitoring campaign of 
NGC~3783 showed the Ly$\alpha$ BLR flux variations lagged the continuum variations by 3.6~days, while 
the \ion{He}{2} BLR flux lag was only 1.3~days \citep{onke02}, indicating the higher ionization
line emission is located closer to the ionizing source. 
If the \ion{O}{6} BLR is also more compact than Ly$\alpha$, then
$C_f^l$ for Ly$\beta$ may be greater than Ly$\alpha$, leading to an overestimate
of the continuum covering factor.  This may explain the discrepancy in Figure~4a, where the continuum covering factor derived at velocities just blueward of the centroid in component~1 and in the core of component~4 is greater than unity.

     Another possible source of error involves the contribution of light from an extended scattering 
region to the residuals in the absorption features.
If the scattered light profile underneath the absorption has structure, this will affect the 
covering factor calculations \citep[e.g., see][]{krae01b}.
This has consequences in using limits on the size of the emission sources
to constrain the size of the absorbers for cases of partial coverage.
However, it seems unlikely that scattered light can explain all of the non-unity covering factors 
derived from the \ion{H}{1} absorption.  For example, to account for all of the residual flux 
in the cores of Ly$\alpha$ in components~1 and 3 would require that 60\% and 50\%, respectively, of the 
intrinsic continuum plus BLR flux is scattered light.
We conclude that, at least in these cases, the low derived covering factors represent 
partial line-of-sight coverage of the emission-line source.  Thus, large variations occur in 
$C_f^l$ between the cores of the different components, with values decreasing for higher 
blueshifted radial velocities in components 1--3.  We note that unobscured NLR emission cannot 
account for the low line covering factors, since the low $C_f^l$ values occur at high outflow 
velocities.  At most, the NLR only affects the red-wing of the lowest velocity component.  
Furthermore, it is difficult to see how the 
scattered light profile could explain the dramatic variation in the continuum covering factor derived in the 
blue wings of components~1 and 3.  In the outer velocities of the wings, the scattered contribution would have to account for 60--70\% of the total intrinsic flux.  
Thus, although we cannot rule out that the continuum source is fully occulted by the absorber
in the cores of all the components, the decreasing
covering factor derived in the {\it wings} of components~1 and 3 implies velocity-dependent, 
partial coverage of the continuum source. 

\subsection{Column Density Measurements}
     For doublet line absorption, the effective covering factor, which is a combination of the
individual line and continuum covering factors, can be derived directly
from the residual fluxes in the two lines using their intrinsic optical depth ratio 
\citep{hama97}.  However, many of the doublets in NGC~3783 are too heavily 
contaminated with other absorption to apply this technique.  In these cases and for lines 
not arising from doublets, another measure of $C_f$ is required that accounts for the individual 
line and continuum covering factors, as highlighted in our analysis of the Lyman lines.  This is 
evident in component~1, in which $C_f =$~0.63 is measured in the core of \ion{N}{5} using the 
doublet method, but \ion{C}{4} and \ion{O}{6} have only one measurable line.  Figure~3 shows the 
\ion{O}{6}, \ion{C}{4}, and Ly$\alpha$ absorption depths are all relatively shallow compared to 
\ion{N}{5} in this component.  Our fits to the continuum and line emission reveal they also 
have a stronger relative contribution of BLR emission at the wavelengths of component~1 absorption 
($R_l =$~0.74, 0.82, and 0.85 for \ion{O}{6}, \ion{C}{4}, and Ly$\alpha$, respectively,
compared to $R_l =$~0.58 for \ion{N}{5}).  In \S~3.1, we found the Ly$\alpha$ absorption is
saturated but it appears weak because of a low covering factor of the BLR. 
Similarly, the shallow absorption observed in \ion{O}{6} and \ion{C}{4} may also be the result of 
relatively low effective covering factors rather than low column densities.  This was suggested by the 
photoionization modeling results of \citet{krae01a}, which showed that substantial 
columns of \ion{C}{4} and/or \ion{O}{6} should always accompany a large \ion{N}{5} column.

     Motivated by these results, we used the covering factors for \ion{H}{1} to
derive effective covering factors and column densities for all intrinsic absorption lines.
Following \citet{gang99}, the effective covering factor is defined in terms of the 
individual line and continuum covering factors, 
\begin{equation}
C_f = R_l C_f^l + R_c C_f^c.
\end{equation}
For each line, we computed $C_f$ as a function of radial velocity from the $C_f^c$ and $C_f^l$ 
profiles derived for \ion{H}{1} and our fits to the intrinsic continuum and line fluxes.  
The effective covering factors, averaged over 50~km~s$^{-1}$ intervals in the cores of the
absorption components, are listed in Table~4.  Uncertainties were derived by propagating the
errors associated with $C_f^c$ and $C_f^l$.
For the doublet absorption in which the cores of both lines are uncontaminated with other 
absorption, i.e., \ion{Si}{4} component~1 and \ion{N}{5} components~1--3,
we also computed the effective covering factors in the cores of the absorption directly using the doublet 
method as described in \citet{hama97}.  These values, and associated uncertainties, are listed in parentheses in Table~4.  These results show the effective covering factor derived for \ion{H}{1} in component~1 is equivalent to that determined for \ion{N}{5}, but is larger than the \ion{Si}{4} covering factor by more than a factor of two.  
{\it This provides the only evidence for a covering factor within an individual kinematic component 
that is a function of ionization.}  We note this may indeed be the case since component~1 
likely consists of at least two distinct subcomponents, one of which is lower ionization and responsible 
for all of the \ion{Si}{4} absorption \citet[see][]{krae01a}.
Comparison of the line widths in this component reveals further evidence of substructure. 
In Table~5, we give the FWHM and centroid velocities measured directly from the normalized absorption 
profiles for each line in component~1.  The \ion{Si}{4} and \ion{N}{3}$^*$ lines are significantly narrower 
than the higher ionization lines, \ion{N}{5} and \ion{C}{4}.  

     Substituting equation~3 into equation~2 gives the familiar expression for normalized flux as a function 
of the effective covering factor \citep{hama97},
\begin{equation}
I = C_f  e^{-\tau} + 1 - C_f. 
\end{equation}
The factor $1 - C_f$ in equation~4 is the normalized unocculted flux.  
This represents the fraction of the total intrinsic flux that is unobscured by the absorber.
The normalized absorption profiles for \ion{O}{6}, \ion{N}{5}, \ion{C}{4}, and \ion{H}{1} are plotted 
together with the normalized unocculted flux levels derived from equation~4 (dashed lines) for component~1 in Figure~5a
and components~2+3 in Figure~5b.  In these figures we have plotted only the least contaminated line for the 
doublets.  Figure~5 illustrates how the apparent absorption is a convolution of covering factor and column 
density.  For lines with large column densities, the normalized flux will approach the unocculted flux level, 
but never lie below it (see equation~4).  Other than the detector artifact in \ion{O}{6}~$\lambda$1038 
component~1 (Figure~5a), we find no cases where residual fluxes lie below the unocculted flux levels outside 
the 1~$\sigma$ uncertainties.  This supports the validity of using the \ion{H}{1} covering factors.   

     In Figure~5, the absorption depths in several lines closely match the derived unocculted flux 
levels, indicating possible saturation in these lines.  For example, the residual flux in the 
\ion{O}{6}~$\lambda$1038 feature in components~2 and 3 traces the unocculted flux levels over nearly the 
entire profile.  The \ion{O}{6}~$\lambda$1032 absorption gives a consistent result in the narrow windows of 
component~2 that are uncontaminated with Galactic absorption ($v_r \approx -$650 and $-$450~km~s$^{-1}$
in Figure~3b).  Additionally, the absorption in the blue wings of \ion{N}{5}~$\lambda$1238, 
\ion{C}{4}~$\lambda$1548, and \ion{O}{6}~$\lambda$1038 in component~1 appears saturated.  However, the 
absorption is seen to diverge from the unocculted flux in the red wing of \ion{N}{5} and \ion{C}{4} (\ion{O}{6} 
is contaminated with a detector artifact at these velocities).  Thus, if the covering factor profiles derived 
for \ion{H}{1} are valid for these lines, the column densities are not uniform across the profile in 
component~1, providing further evidence for substructure in this component.  Given the uncertainties in the 
covering factor profiles due to the uncertain scattered light profile and the possibility that the covering 
factors of the \ion{O}{6} and Ly$\alpha$ BLRs are not the same, as described in \S~3.1, we cannot rule out that this apparent substructure is due to our assumptions about the covering factor profiles.

     Equation~4 can be rearranged to give the optical depth \citep{hama97},
\begin{equation}
\tau_{\rm{v}} = \ln(\frac{C_f}{I - 1 + C_f}),
\end{equation}
where the subscript indicates the calculation is made at each radial velocity, thereby
giving the optical depth profile.  The column density at each radial velocity is then obtained 
from the optical depth,
\begin{equation}
N_{\rm{v}}=\frac{m_e c}{\pi e^2 f \lambda} \tau_{\rm{v}},
\end{equation}
\citep{sava91}.  We measured the total column density for each line by integrating 
equation~6 over the optical depth profiles calculated from equation~5; results are given in
Table~6.  For the doublets where a direct measurement of the effective covering factor is available from the 
residuals in the two lines, we used the core value listed in parenthesis in Table~4 to compute 
the column density.  Components 2 and 3 were deblended following the method outlined in \citet{krae01a}.
In saturated regions, we measured lower limits for the column densities using 
$I+\Delta I$ in equation~5, where $\Delta I$ is the 1~$\sigma$ uncertainty in the normalized flux. 
Upper limits on undetectable lines were determined from uncertainties in the normalized fluxes.  
The 1~$\sigma$ uncertainties quoted in Table~6 were estimated from repeated measurements, and include 
uncertainties in our fits to the intrinsic emission as well as the photon noise.  
We note that the lower limits on the \ion{O}{6} columns from the combined components are 
consistent with the upper limit of 10$^{17}$~cm$^{-2}$ derived from the X-ray spectrum in Paper~I.
If the \ion{P}{5} absorption in component~1 is real (see discussion in \S 2), then it has a column
density of 1.5~$\pm$~0.7~$\times$~10$^{13}$~cm$^{-2}$.

    Given the important implications of our results involving the covering 
factors, particularly the lower covering factor derived for \ion{Si}{4}, it is worthwhile to address the 
possible systematic errors that were discussed in \S 3.1 in more depth.
First, we note that any contribution of scattered light to the residual fluxes in the 
Ly$\beta$ or Ly$\alpha$ absorption will lead to an {\it underestimate} of the 
covering factors.  This would lead to an even larger discrepancy 
between the effective covering factor derived for \ion{Si}{4} from the doublet and the
prediction from the \ion{H}{1} results. 
The second possible systematic error is introduced if the sizes 
of the \ion{O}{6} and Ly$\alpha$ BLRs are not identical, leading to different emission
line covering factors for Ly$\beta$ and Ly$\alpha$.
Since the Ly$\epsilon$ absorption has no underlying BLR emission (see Table~3), it can
be used to place limits on the covering factors that are unaffected by
any discrepancy in the sizes of the BLRs of the different lines.
For example, in component~1, an absolute lower limit on the \ion{H}{1} {\it continuum}
covering factor in the core follows directly from the depth of the feature,
$C_f^c >$~1$-I=$0.77.
This is stricly a lower limit, since (a) Ly$\epsilon$ is not saturated and (b)
there may be a contribution from scattered flux.
Using this result in equation~3, we derive an absolute lower limit on the effective covering 
factor predicted for \ion{Si}{4} from the \ion{H}{1} covering factors, $C_f >$~0.58.
Comparing this with the value derived from the \ion{Si}{4} doublet, 
$C_f =$~0.30~$\pm$0.07, we conclude that our finding that the covering factor of \ion{Si}{4} 
is less than \ion{H}{1} is robust.

\section{Constraints on the Physical Conditions and Geometry of the Absorbers}
     While detailed photoionization modeling of the absorption will be presented in future
papers (Kraemer et al.\ in preparation; Netzer et al.\ in preparation), we can place some
immediate constraints on the physical conditions and geometry of the low ionization UV
absorber in component~1.  We refer to the photoionization modeling results in \citet{krae01a}, 
which predicted a total hydrogen column density, $N_H =$~5~$\times$~10$^{18}$~cm$^{-2}$, and ionization 
parameter, $U =$~0.0018, for this subcomponent.  Since the \ion{Si}{4} column measured in the 
averaged spectrum is somewhat higher than in the \citet{krae01a} study, partly due to 
the lower covering factor found for this line in the present study, and we have a more rigid upper limit on the 
\ion{C}{2} column because of the high S/N, we have modified the models slightly.
Using the ionizing spectral energy distribution and abundances employed in the models of
\citet{krae01a}, we find that $N_H \geq$~10$^{19}$~cm$^{-2}$ and $U \geq$~0.005 are 
required to match the \ion{Si}{4} column and \ion{C}{2} upper limit simultaneously.

     The detection of \ion{C}{3}$^* \lambda$1175--76 absorption, which arises from metastable 
levels 6.5~eV above the ground state, implies a high density in the low ionization absorber
in component~1 \citep{brom85,kris92,krae01b}.  
The relative population of the metastable levels is determined by the physical conditions
in the absorber, and the ratio of metastable to ground state columns can be used to estimate the 
electron density, $n_e$ \citep{kris92}.  Due to contamination of the 
\ion{C}{3}~$\lambda$977 feature by Galactic H$_2$, the \ion{C}{3} ground state column density 
cannot be measured directly in NGC~3783.  Instead, we use the photoionization 
model results, which predict $N_{C III} \approx$~2~$\times$~10$^{15}$~cm$^{-2}$.  The predicted
\ion{C}{3} column is not a strong function of the exact $U$, $N_H$ solution needed to match the \ion{Si}{4}, 
since \ion{Si}{4} and \ion{C}{3} appear in the same zone due to the similarity in their ionization 
potentials.  Our measured column for the metastable level from \ion{C}{3}$^* \lambda$1175--76    
(1.2~$\times$~10$^{13}$~cm$^{-2}$) then implies the ratio of metastable to ground state columns is $\sim$0.006. 
The photoionization models predict $T \approx$~20,000~K in the absorber, which is largely insensitive
to the exact $N_H$, $U$ solution for our assumed SED.  Using the calculations in \citet{kris92}, this corresponds to 
$n_e \approx$~10$^9$~cm$^{-3}$ for the low ionization subcomponent.  
This high density implies a very small radial dimension for the absorber, $\sim$10$^{10}$~cm for a column
of 10$^{19}$~cm$^{-2}$, which may indicate the gas is clumpy and has a low volume filling factor
(see discussion below).  From the density, lower limit on $U$, and estimate of the ionizing continuum luminosity 
given by \citet{krae01a}, we derive an upper limit on the location of the absorber with respect to the ionizing continuum source, $R \leq$~8~$\times$~10$^{17}$~cm.

     The partial coverage of the emission sources revealed in the analysis of the \ion{H}{1} lines
in \S~3.1 also places constraints on the absorption geometry.  
At velocities exhibiting partial coverage, the area of the absorbing gas projected
on the sky is smaller than the projected area of the emission sources.
The monitoring study of NGC~3783 revealed large amplitude variations of both the continuum and BLR emission 
\citep{reic94,onke02}.
The most rigid constraint comes from the measured lag of the BLR flux variations behind the continuum
variations. For example, \citet{onke02} found the Ly$\alpha$ BLR flux varied by up to a factor of 1.6 and lagged the continuum variations by 3.6~days.  From light travel time arguments, this constrains the projected size of much of the BLR to be $\sim$3.6~lt-days (10$^{16}$~cm) across.  Thus, in components~1 and 3, with $C_f^l =$~0.31 and 0.55 measured in the cores, the projected area of the absorbers is $\sim$(5.5~$\times$~10$^{15}$)$^2$ and $\sim$(7.5~$\times$~10$^{16}$)$^2$~cm$^2$, respectively.

     These results provide important clues to the physical nature of the absorbers.  
The smoothly varying covering factor of the continuum detected in the wings
of the absorption in \S 3.1 indicates there is substructure within the individual
kinematic components.  One interpretation of this result is that the absorption
components are comprised of many small substructures, such as cloudlets or turbulent
cells.  In this scenario, the decreasing coverage in the wings may be a consequence
of the velocity dispersion of these substructures, with a decreasing number of ``clouds" having velocities
that diverge from the centroid velocity.  Since, in general, the ionic column densities do not decrease 
dramatically in the wings of the absorption profiles (see Figure~5), there cannot be much overlap of clouds 
in the cores of the absorbers, indicating they have small volume filling factors.
The extreme compactness of the gas comprising the low ionization subcomponent in 
component~1 provides additional evidence for this model. 
Since the radial thickness of this gas is much smaller than the projected 
dimensions of the emission sources, the most natural explanation is that a large
number of very small, dense clouds combine to give the observed absorption.
The alternative interpretation would require that the absorber is an exceedingly thin sheet of material.
Previous studies have invoked absorption by a clumpy gas to explain the complex structure seen 
in other AGN absorption systems \citep[e.g.,][]{hama01,kris02} and a highly inhomogeneous absorbing 
medium is predicted by the multitemperature wind models of \citet{krol01}.

\section{Summary}

     NGC~3783 was observed with {\it HST}/STIS at 18 epochs and with
{\it FUSE} at 5 epochs as part of an intensive multiwavelength monitoring campaign designed 
to investigate the intrinsic absorption in this Seyfert~1 galaxy.
We have combined the observations to produce a high S/N averaged spectrum in the UV and far-UV.
The major findings of our study follow.

1.  \ion{O}{6}, \ion{N}{5}, \ion{C}{4}, \ion{N}{3}, and the Lyman lines up to
Ly$\epsilon$ appear in the three kinematic components identified by \citet{krae01a} in the 
initial STIS spectrum (components 1, 2, and 3 at radial velocities $-$1320, $-$548, and $-$724~km~s$^{-1}$, respectively) .  Galactic contamination prevents the detection of higher order Lyman lines.
\ion{Si}{4} appears in component~1; \ion{C}{3}~$\lambda$977
is heavily contaminated with Galactic absorption and cannot be measured in this component, but is
present in components 2 and 3.  
We detect absorption from the \ion{C}{3}$^* \lambda$1175-76 multiplet in component~1,
indicating a high density in this absorber.  This is the first detection of metastable
\ion{C}{3} in a Seyfert galaxy other than NGC~4151 \citep{brom85}.
No lower ionization lines appear in absorption, and we place stringent upper limits on their 
column densities.  A fourth kinematic component of intrinsic absorption is tentatively identified 
at a radial velocity of $-$1027~km~s$^{-1}$.  This component appears strong only in Ly$\beta$, which
may be contaminated with Galactic absorption, and \ion{O}{6}.  In all other lines, the detection
of this component is very sensitive to the fit to the intrinsic emission.

2.  The Lyman lines reveal a complex absorption geometry in NGC~3783 and highlight the importance
of determining the individual covering factors of the continuum and emission-line sources.  
The strength of the higher order Lyman lines indicates Ly$\alpha$ and Ly$\beta$ are saturated in components 1--3.
The continuum and emission-line covering factors were separated using the Ly$\alpha$ and Ly$\beta$ 
absorption in each kinematic component, both in the cores of absorption and as a function of radial velocity.  The covering factor of the BLR varies significantly between the cores of the absorption components, with $C_f^l =$~0.33, 0.84, 0.55, and 0.03 derived in components 1, 2, 3, and 4, respectively. 
The large residual fluxes measured in Ly$\alpha$ in components~1 and 3 imply that, at least in these components, the non-unity emission-line covering factors cannot be fully accounted for by light from an extended scatterer, and thus represent partial coverage of the BLR.  
We also find evidence for variation of the continuum covering factor with velocity in the individual kinematic components.  Specifically, $C_f^c$ decreases by $\sim$60\% over several resolution elements in the wings of the absorption.  This smoothly decreasing coverage is a signature of substructure within the absorbers.

3.  The individual continuum and line covering factors derived from \ion{H}{1} were used to derive
effective covering factors for all lines.
Comparison to the covering factors derived directly by the doublet method
reveals the \ion{Si}{4} covering factor is smaller than that of \ion{H}{1} and \ion{N}{5} by
more than a factor of two in component~1.  
This implies substructure in component~1 and is consistent with the prediction by \citet{krae01a} that 
this component is comprised of at least two zones of UV absorption.  Furthermore, the FWHM
of \ion{Si}{4} and \ion{N}{3} are narrower than the higher ionization lines in component~1, providing 
additional evidence for substructure.
Employing the \ion{H}{1} covering factor profiles for the higher ionization lines, we find
the relatively weak apparent \ion{C}{4} and \ion{O}{6} absorption 
compared to \ion{N}{5} in component~1 is due to low effective covering factors rather than  
small column densities.
If the covering factors derived from \ion{H}{1} are valid for \ion{C}{4} and \ion{N}{5}, then the column 
densities of these lines are not uniform across the profile of component~1;
they are saturated in the blue wing, but smaller in the red wing.
Additionally, \ion{O}{6} is found to be saturated in components~2 and 3, as well as in the blue
wing of component~1.

4.  We place lower limits on the total hydrogen column density and ionization 
parameter in the low ionization subcomponent in component~1, $N_H \geq$~10$^{19}$~cm$^{-2}$ and
$U \geq$~0.005.  Combining model predictions of the total \ion{C}{3} column with our measured 
\ion{C}{3}$^*$ column gives $n_e \approx$~10$^{9}$~cm$^{-3}$.  These results imply an upper limit
on the distance of the absorber from the ionizing continuum source of $\leq$~8~$\times$~10$^{17}$~cm.
Using upper limits on the size of the BLR from the variability study by 
\citet{onke02}, the projected area of the absorbers at velocities exhibiting
partial coverage of the emission-line source is found to be $\sim$(10$^{16}$)$^2$~cm$^2$.
The decreasing covering factor exhibited in the wings of the Lyman lines and extreme compactness
of the \ion{C}{3}$^*$ absorber are suggestive of a clumpy absorbing medium with a low filling factor.

     J. R. G., D. M. C., and S. B. K. acknowledge support from NASA grant HST-GO-08606.13-A.  W.N.B
acknowledges CXC grant GO1-2103 and NASA LTSA grant NAG5-8107.  F.H. acknowledges NSF grant AST 99-84040.
We thank B-G Andersson and the {\it FUSE} team for assistance with the {\it FUSE} spectra and Derck 
Massa for providing his code for modeling the H$_2$ absorption.  We also thank the anonymous referee 
for very helpful comments.

\clearpage

\begin{deluxetable}{lrrr}
\tablewidth{0pt}
\tablecaption{{\it HST}/STIS Observations of NGC 3783\tablenotemark{a} \label{tbl-1}}
\tablehead{
\colhead{Data Set} & \colhead{UT Start Time} & \colhead{Exp Time} \\
\colhead{} & \colhead{} & \colhead{(seconds)}}
\startdata
O57B01020\tablenotemark{b} &  2000-02-27, 09:34:59 &   2700   \\
O57B01030\tablenotemark{b} &  2000-02-27, 11:11:33 &   2700   \\
O63M01010 &  2000-08-05, 00:54:58 &   2200   \\
O63M01020 &  2000-08-05, 02:18:41 &   2700   \\
O63M02010 &  2000-11-26, 01:10:17 &   2200   \\
O63M02020 &  2000-11-26, 02:35:18 &   2700   \\
O63M03010 &  2001-01-25, 20:09:07 &   2200   \\
O63M03020 &  2001-01-25, 21:34:05 &   2700   \\
O63M04010 &  2001-02-25, 04:01:51 &   2200   \\
O63M04020 &  2001-02-25, 05:22:57 &   2700   \\
O63M05010 &  2001-02-28, 01:05:49 &   2200   \\
O63M05020 &  2001-02-28, 02:28:43 &   2700   \\
O63M06010 &  2001-03-02, 17:20:14 &   2200   \\
O63M06020 &  2001-03-02, 18:43:59 &   2700   \\
O63M08010 &  2001-03-11, 14:54:01 &   2200   \\
O63M08020 &  2001-03-11, 16:20:48 &   2700   \\
O63M09010 &  2001-03-15, 16:51:54 &   2200   \\
O63M09020 &  2001-03-15, 18:19:14 &   2700   \\
O63M10010 &  2001-03-19, 17:12:53 &   2200   \\
O63M10020 &  2001-03-19, 18:40:08 &   2700   \\
O63M11010 &  2001-03-23, 15:59:27 &   2200   \\
O63M11020 &  2001-03-23, 17:24:31 &   2700   \\
O63M12010 &  2001-03-27, 08:19:43 &   2200   \\
O63M12020 &  2001-03-27, 09:43:10 &   2700   \\
O63M14010 &  2001-04-04, 02:35:42 &   2200   \\
O63M14020 &  2001-04-04, 03:57:29 &   2700   \\
O63M15010 &  2001-04-08, 01:22:01 &   2200   \\
O63M15020 &  2001-04-08, 02:41:30 &   2700   \\
O63M16010 &  2001-04-11, 22:33:08 &   2200   \\
O63M16020 &  2001-04-11, 23:49:35 &   2700   \\
O63M07010 &  2001-04-16, 02:09:01 &   2242   \\
O63M07020 &  2001-04-16, 03:28:27 &   2700   \\
O63M53010 &  2001-04-23, 15:37:54 &   2242   \\
O63M53020 &  2001-04-23, 17:02:53 &   2700   \\
O63M17010 &  2002-01-06, 13:39:42 &   2242   \\
O63M17020 &  2002-01-06, 15:15:51 &   2700   \\
\tablenotetext{a}{All data obtained with the E140M echelle grating, through the 
0$\farcs$2~$\times$~0$\farcs$2 aperture.}
\tablenotetext{b}{Data for this observation were presented in Kraemer et al.\ (2001a).}
\enddata
\end{deluxetable}
\clearpage

\begin{deluxetable}{lrrr}
\tablewidth{0pt}
\tablecaption{{\it FUSE} Observations of NGC 3783\tablenotemark{a} \label{tbl-2}}
\tablehead{
\colhead{Data Set} & \colhead{UT Start Time} & \colhead{Exp Time} \\
\colhead{} & \colhead{} & \colhead{(seconds)}}
\startdata
B1070102000  &  2001-02-28, 16:46 & 27830 \\
B1070106000  &  2001-03-07, 08:31  & 26400 \\
B1070103000  &  2001-03-11, 12:56  & 27494   \\
B1070105000  &  2001-03-30, 22:52   & 28099 \\
B1070105000  &  2001-06-27, 05:26   & 27490 \\
\tablenotetext{a}{All data obtained through the 30$\arcsec$~$\times$~30$\arcsec$ LWRS aperture.}
\enddata
\end{deluxetable}
\clearpage
    
\begin{deluxetable}{lrrrrrrrrrrr}
\tablewidth{0pt}
\tablecaption{Continuum and Emission-Line Covering Factors in the Cores of the Lyman Lines \label{tbl-3}}
\tablehead{
\colhead{Comp} & \colhead{} & \colhead{} & \colhead{$I$\tablenotemark{a}} & \colhead{} & \colhead{} &  
\colhead{} & \colhead{} & \colhead{$R_c$\tablenotemark{b}} & \colhead{} &  \colhead{$C_f^c$\tablenotemark{c}} &
\colhead{$C_f^l$\tablenotemark{c}}\\
\colhead{}  &\colhead{Ly$\alpha$} & \colhead{Ly$\beta$} & \colhead{Ly$\gamma$} & \colhead{Ly$\epsilon$} & 
\colhead{} & \colhead{Ly$\alpha$} & \colhead{Ly$\beta$} & \colhead{Ly$\gamma$} & \colhead{Ly$\epsilon$} & \colhead{} & \colhead{}}
\startdata
1  & 0.57  & 0.27  & \nodata  & 0.23  &  &  0.15 & 0.57 & 0.92 & 1.0 & 1.0 ($\pm$0.10) & 0.31 ($\pm$0.06) \\
2  & 0.18  & 0.20  & 0.38  & 0.50  &  &  0.10 & 0.52 & 0.89 & 1.0 & 0.74 ($\pm$0.08) & 0.84 ($\pm$0.08)\\
3  & 0.41  & 0.26  & 0.24  & 0.46  &  &  0.11 & 0.52 & 0.90 & 1.0 & 0.9 ($\pm$0.10) & 0.55 ($\pm$0.08) \\
\tablenotetext{a}{Residual normalized flux in core measured over $\Delta v = -$1325 -- $-$1275,
$-$575 -- $-$525, $-$750 -- $-$700, and $-$1050 -- $-$1000 km~s$^{-1}$ for components 1, 2, 3, and 4 respectively.}
\tablenotetext{b}{Fractional contribution of continuum to total intrinsic emission.}
\tablenotetext{c}{Covering factors of the continuum, $C_f^c$, and line, $C_f^l$, emission sources; quoted 
uncertainties do not include possible errors due to a contribution from scattered flux or structure in
the BLR.}
\enddata
\end{deluxetable}
\clearpage

\begin{deluxetable}{lllll}
\tablecaption{Effective Covering Factors Derived in the Cores of Absorption Lines 
using \ion{H}{1}\tablenotemark{a}\label{tbl-4}}
\tablewidth{0pt}
\tablehead{
\colhead{Line} & \colhead{Comp 1} & \colhead{Comp 2}  & \colhead{Comp 3}  & \colhead{Comp 4}}
\startdata
\ion{O}{6} $\lambda$1038  & 0.52$\pm$0.07 & 0.81$\pm$0.08 & 0.66$\pm$0.09 & 0.29 \\
\ion{N}{5} $\lambda$1238  & 0.63$\pm$0.07 (0.63$\pm$0.04) & 0.81$\pm$0.08 (0.61$\pm$0.05) & 0.70$\pm$0.09 (0.66$\pm$0.04) &  0.41 \\
\ion{N}{3}* $\lambda$991 & 0.93$\pm$0.10 & 0.78$\pm$0.08 & 0.86$\pm$0.10 & \nodata \\
\ion{C}{4} $\lambda$1548 & 0.45$\pm$0.07 & 0.82$\pm$0.08 & 0.60$\pm$0.08 & 0.19 \\
\ion{C}{3} $\lambda$977 & \nodata & 0.79$\pm$0.08 & 0.88$\pm$0.10 & 0.84 \\
\ion{C}{3}* $\lambda$1775-76\tablenotemark{c} & 0.88$\pm$0.10 & \nodata & \nodata & \nodata \\
\ion{Si}{4} $\lambda$1393 & 0.72$\pm$0.07 (0.30$\pm$0.07) & \nodata & \nodata & \nodata \\
\enddata
\tablenotetext{a}{Covering factors derived using $C_f^l$ and $C_f^c$ derived for \ion{H}{1}; effective covering
factors derived directly from doublet method are given in parenthesis.  Values are averages measured over
$\Delta v = -$1325 -- $-$1275, $-$575 -- $-$525, $-$750 -- $-$700, and $-$1050 -- $-$1000 km~s$^{-1}$ for 
components 1, 2, 3, and 4 respectively.}
\end{deluxetable}
\clearpage

\begin{deluxetable}{lrr}
\tablecaption{Kinematics in Absorption Component 1 \label{tbl-5}}
\tablewidth{0pt}
\tablehead{
\colhead{Line} & \colhead{Velocity\tablenotemark{a}} & \colhead{FWHM}\\ 
\colhead{} & \colhead{km s$^{-1}$} & \colhead{km s$^{-1}$}}
\startdata
\ion{N}{5} $\lambda$1238  & $-$1304 ($\pm$10) & 194 ($\pm$7) \\
\ion{C}{4} $\lambda$1548 & $-$1322 ($\pm$10) & 180 ($\pm$7) \\
\ion{Si}{4} $\lambda$1393 & $-$1326 ($\pm$15) & 111 ($\pm$10) \\
\ion{N}{3}* $\lambda$991  & $-$1331 ($\pm$20) & 124 ($\pm$15) \\
Ly$\beta$ & $-$1297 ($\pm$12) & 219 ($\pm$10) \\
Ly$\epsilon$ & $-$1315 ($\pm$20) & 145 ($\pm$15) \\
\enddata
\tablenotetext{a}{Centroid velocity with respect to the systemic redshift of the host galaxy.}
\end{deluxetable}
\clearpage

\begin{deluxetable}{lllll}
\tablecaption{Ionic Column Densities\tablenotemark{a} ~(10$^{14}$ cm$^{-2}$) \label{tbl-6}}
\tablewidth{0pt}
\tablehead{
\colhead{Ion} & \colhead{Component 1} & \colhead{Component 2}  & \colhead{Component 3}  & \colhead{Component 4}}
\startdata
\ion{H}{1} & 140 ($\pm$20) & 37 ($\pm$9) & 163 ($\pm$30) & 5.0 ($\pm$2.0) \\
\ion{O}{6}  & $>$14 & $>$12 & $>$51 & 6.1 ($\pm$4.0) \\
\ion{N}{5}  & $>$11 & 2.6 ($\pm$0.5) & 11.4 ($\pm$2.0) &  0.50 ($\pm$0.4) \\
\ion{N}{3}  & 4.4 ($\pm$2.5) & \nodata & \nodata  & $<$1.0 \\
\ion{N}{3}*\tablenotemark{b}  & 4.5 ($\pm$1.5) & 0.4 ($\pm$0.2) & 1.5 ($\pm$0.8) & $<$1.0 \\
\ion{C}{4}  & $>$4.5 & 0.60 ($\pm$0.13) & 2.6 ($\pm$0.5) & 0.34 ($\pm$0.25) \\
\ion{C}{3} & \nodata & 0.30 ($\pm$0.2) & 0.90 ($\pm$0.35) &  \nodata \\
\ion{C}{3}*\tablenotemark{c} & 0.12 ($\pm$0.06) & $<$0.05 & $<$0.03 &  $<$0.03 \\
\ion{C}{2} & $<$0.6 & $<$0.25 & $<$0.10 & $<$0.10 \\
\ion{Si}{4} & 0.89 ($\pm$0.20) & $<$0.1 & $<$0.03 & $<$0.03 \\
\ion{Si}{2} & $<$0.06 & $<$0.04 & $<$0.02 & $<$0.01 \\
\enddata
\tablenotetext{a}{Covering factors derived for \ion{H}{1} were used in all measurements,
except where direct measurement of $C_f$ using the doublet method was possible; 
\nodata denotes no measurement available due to blending with other features; lower
limits given for saturated lines.}
\tablenotetext{b}{Population of N$^{+1}$ J=3/2 fine-structure level.}
\tablenotetext{c}{Population of C$^{+2}$ $^3$P$^o$ metastable levels (6.5 eV).}
\end{deluxetable}

\clearpage
\begin{figure}
\epsscale{0.7}
\plotone{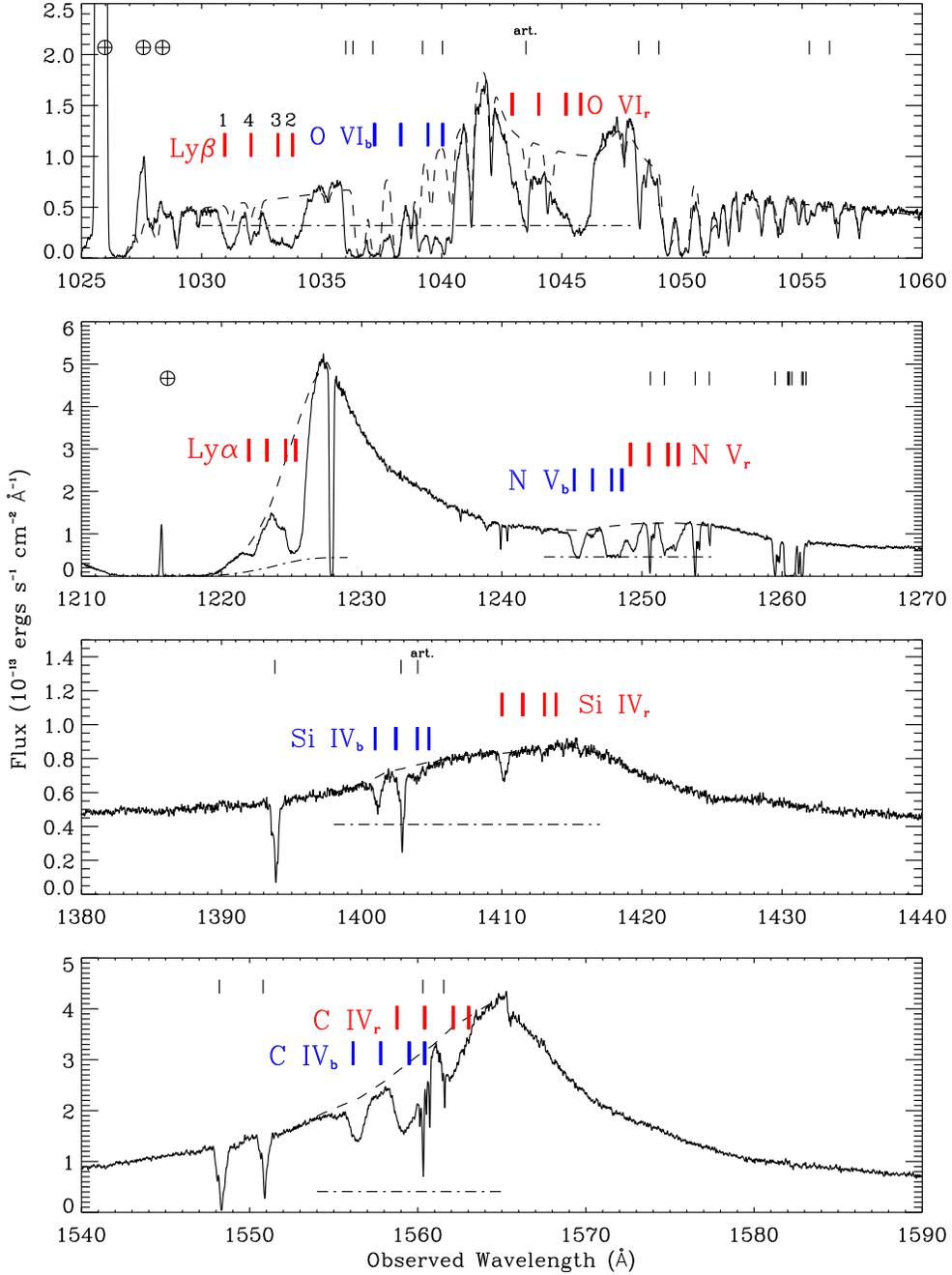}
\caption{Portions of the averaged STIS and {\it FUSE} spectrum of NGC 3783
showing intrinsic absorption.  
Wavelengths corresponding to the four detected absorption components 
are identified with red and blue tick marks.  Labels showing our numbering of the 
components are given above Ly$\beta$.  Fits to the intrinsic
continuum plus line emission are plotted over each line (dashed).
Continuum flux levels are plotted as dot-dashed lines.  
The fit to the total intrinsic flux for \ion{O}{6}, Ly$\beta$ and Ly$\alpha$ 
includes our model of the Galactic H$_2$ and \ion{H}{1} absorption. 
Other strong interstellar lines and detector artifacts (labeled with "art.") 
are denoted with short tick marks at the top of each spectrum.  
Geocoronal emission lines are labeled with $\earth$ symbols. \label{fig1}}
\end{figure}

\clearpage
\begin{figure}
\epsscale{0.7}
\plotone{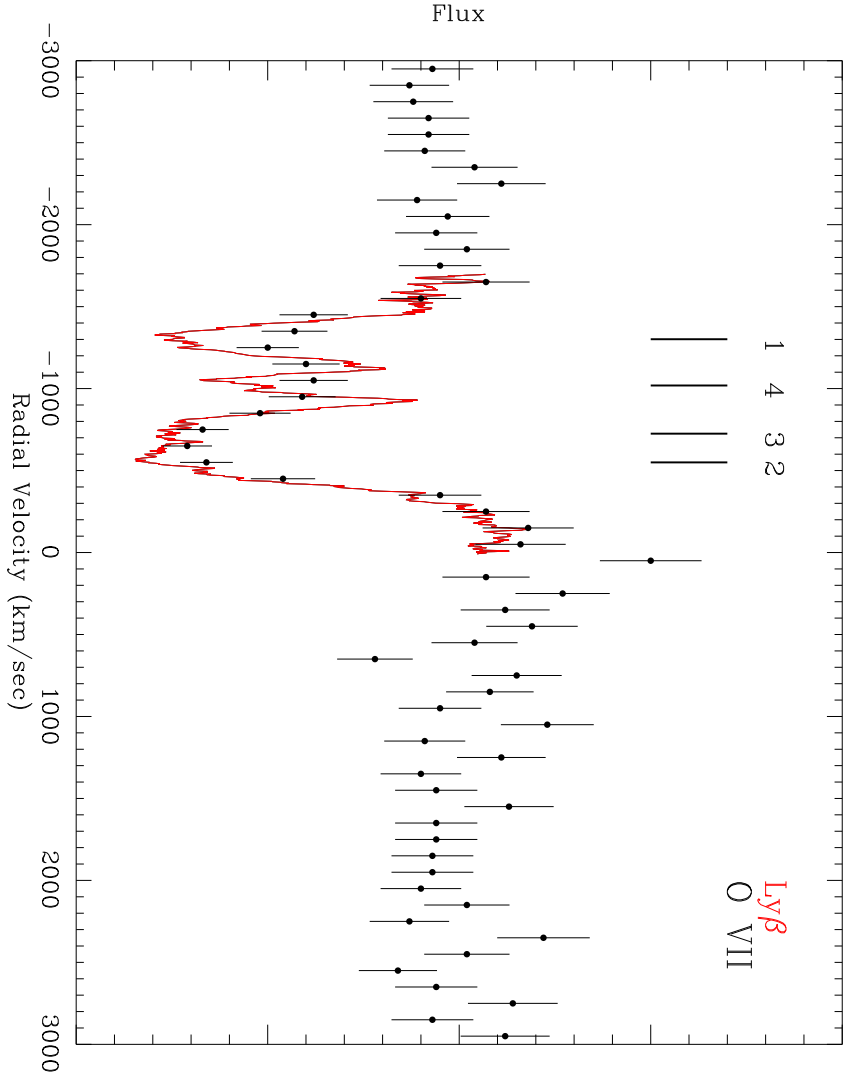}
\caption{Spectra of Ly$\beta$ and \ion{O}{7} showing the kinematic link between
the UV and X-ray absorption in NGC 3783.  The data are plotted in arbitrary flux units. 
The Ly$\beta$ data (red line) are from the averaged {\it FUSE} spectrum.  The \ion{O}{7} profile is a 
combination of the four strongest lines from that ion in the averaged {\it Chandra} spectrum 
(see \S 3.5 in Paper~I); these data are binned at 100 km s$^{-1}$.  Each line is plotted as a function of radial
velocity with respect to the systemic redshift.  The velocities of the four kinematic components
are marked with tick marks above the spectra.  As described in Paper~I, structure is apparent in 
the \ion{O}{7} profile, coinciding in velocity with the UV components. \label{fig2}}
\end{figure}

\clearpage
\begin{figure}
\epsscale{0.5}
\plotone{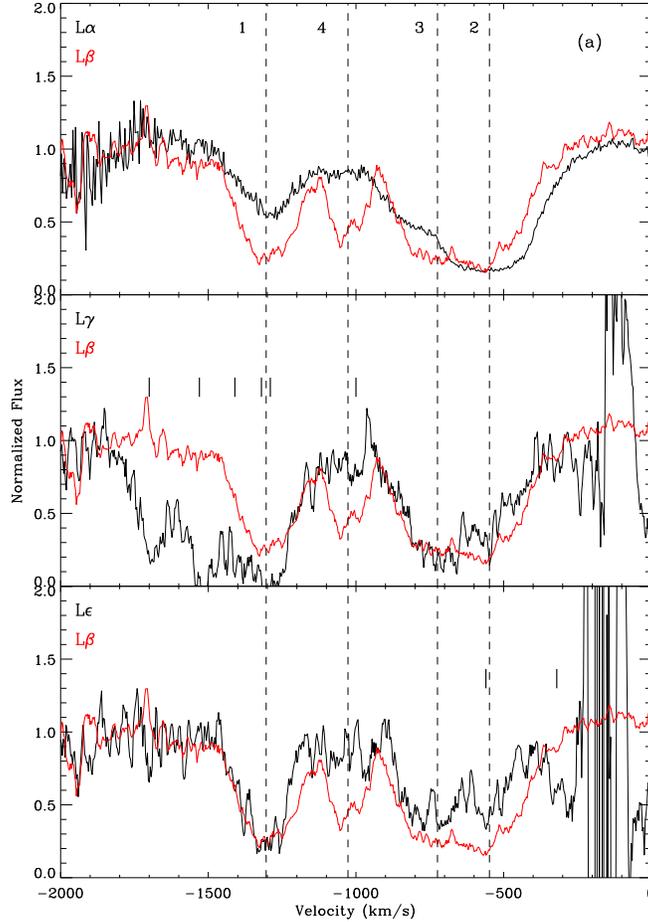}
\caption{Normalized profiles of lines exhibiting intrinsic absorption in
NGC 3783.  The spectra are plotted as a function of radial velocity with
respect to the systemic redshift.  Centroid velocities of the four kinematic absorption components 
are marked with dashed vertical lines.  These were derived from the \ion{N}{5}~$\lambda$1239
profile for components~1--3, and Ly$\beta$ for component~4.
Short tick marks identify strong interstellar lines and detector artifacts (labeled with "art.") 
in each spectrum.  
In 3a, Ly$\beta$ (red line) is plotted together with Ly$\alpha$, Ly$\gamma$, and Ly$\epsilon$.  
In 3b, the short wavelength (blue line) and long wavelength (red line) members of the doublets 
are plotted together.  The tick marks identifying Galactic lines in 3b are color coded to the
corresponding line of the doublet.
In the second panel in 3c, the \ion{C}{3}$^*$ spectrum is plotted with 
respect to the shortest wavelength line in the multiplet.  The positions of
the six multiplet lines for component~1 are marked with long tick marks at the top of the spectrum.
In the \ion{N}{3} $\lambda$989 spectrum (third panel in 3c), component 1 \ion{N}{3}$^* \lambda$991 
absorption is marked with a thick tick mark and labeled. 
Tentative absorption is detected in \ion{P}{5}~$\lambda$1118 in component 1 (bottom
panel in 3c), however this cannot be confirmed in the \ion{P}{5}~$\lambda$1128 line  
due to contamination by Galactic absorption.
All profiles have been corrected for Galactic H$_2$ and \ion{H}{1} absorption.  Regions with 
large fluctuations in the normalized flux are where saturated interstellar lines were divided out.  
\label{fig3} }
\end{figure}

\addtocounter{figure}{-1}

\clearpage
\begin{figure}
\epsscale{0.7}
\plotone{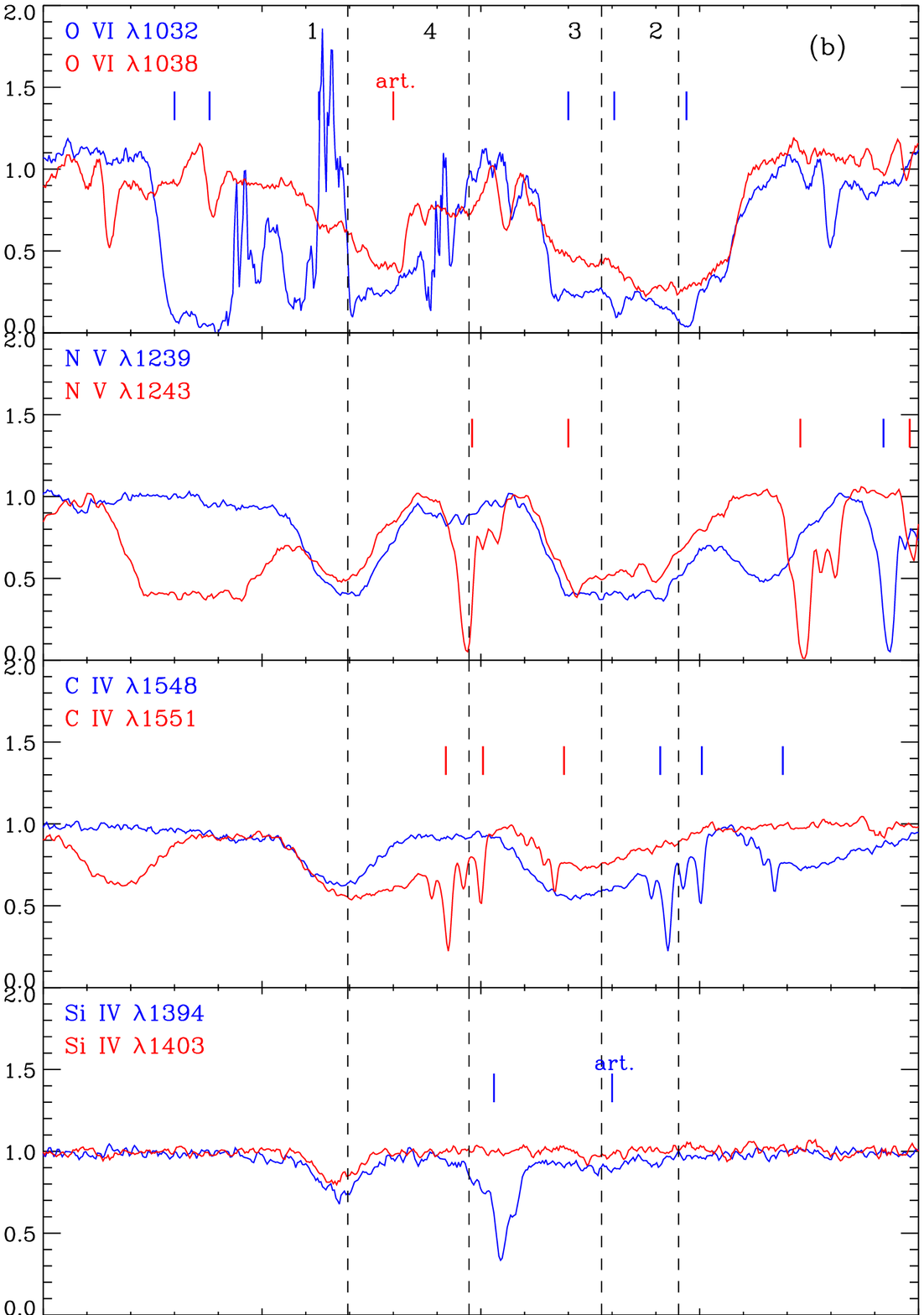}
\caption{Continued \label{fig3b}}
\end{figure}

\addtocounter{figure}{-1}

\clearpage
\begin{figure}
\epsscale{0.7}
\plotone{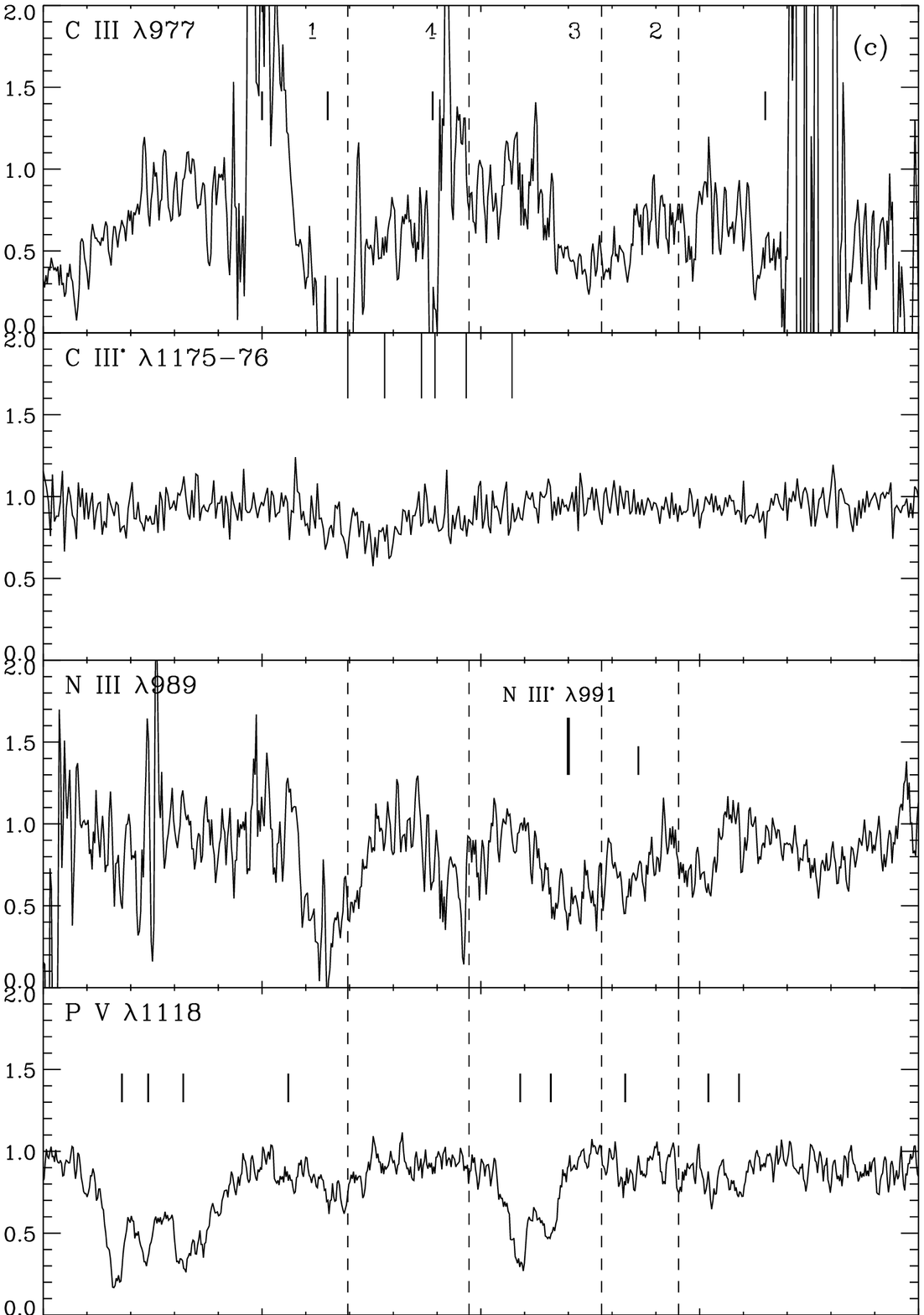}
\caption{Continued \label{fig3c}}
\end{figure}

\clearpage
\begin{figure}
\epsscale{0.7}
\plotone{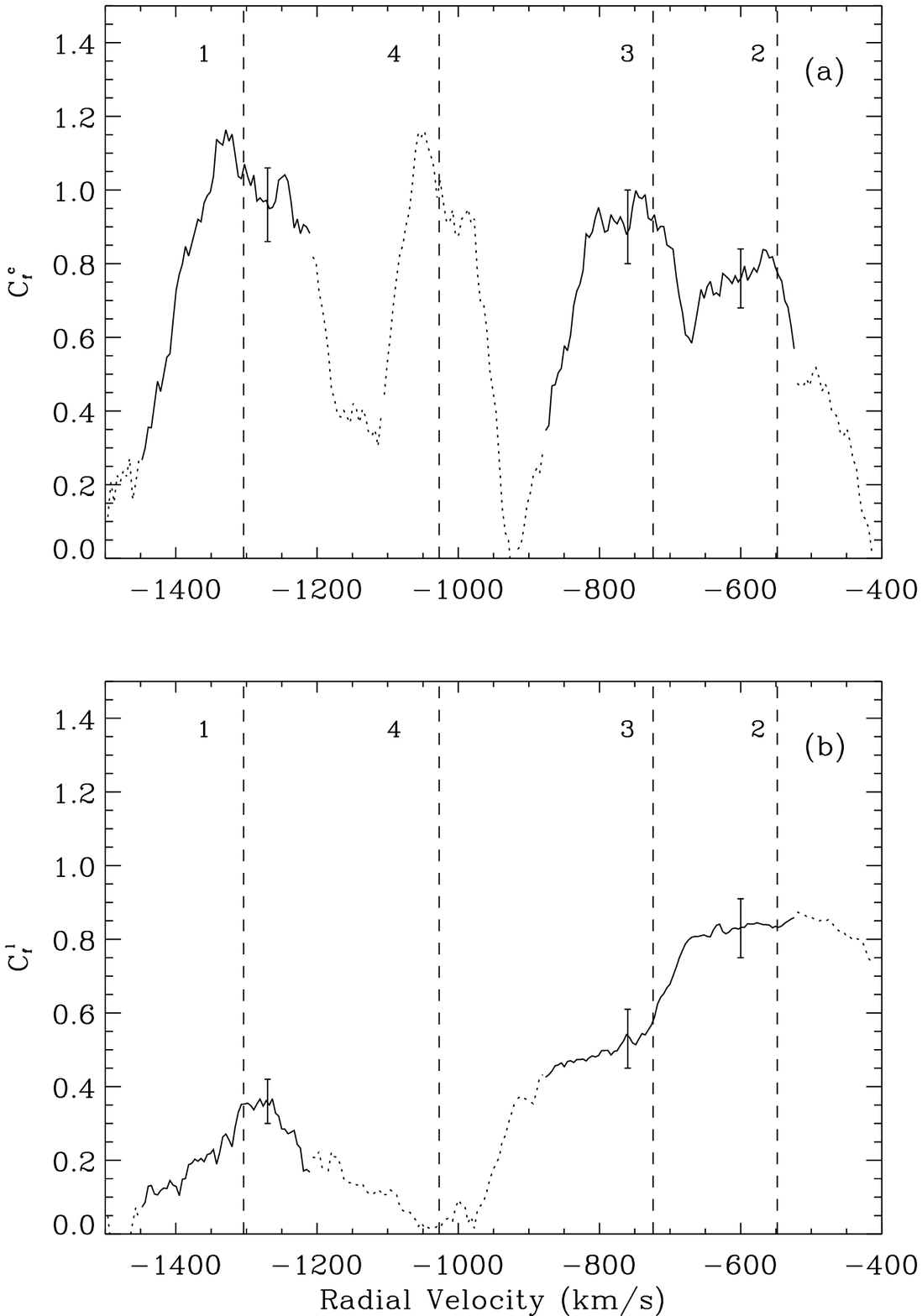}
\caption{Covering factor as a function of radial velocity derived for \ion{H}{1} for the
continuum (4a) and emission-line region (4b).  The dotted line represents lower
limits to $C_f^c$ and $C_f^l$, where Ly$\beta$ is not saturated.  Centroid velocities of 
the four kinematic absorption components are marked with dashed vertical lines.  Sample
error bars are included in the cores of components 1--3.\label{fig4}}
\end{figure}

\clearpage
\begin{figure}
\epsscale{0.7}
\plotone{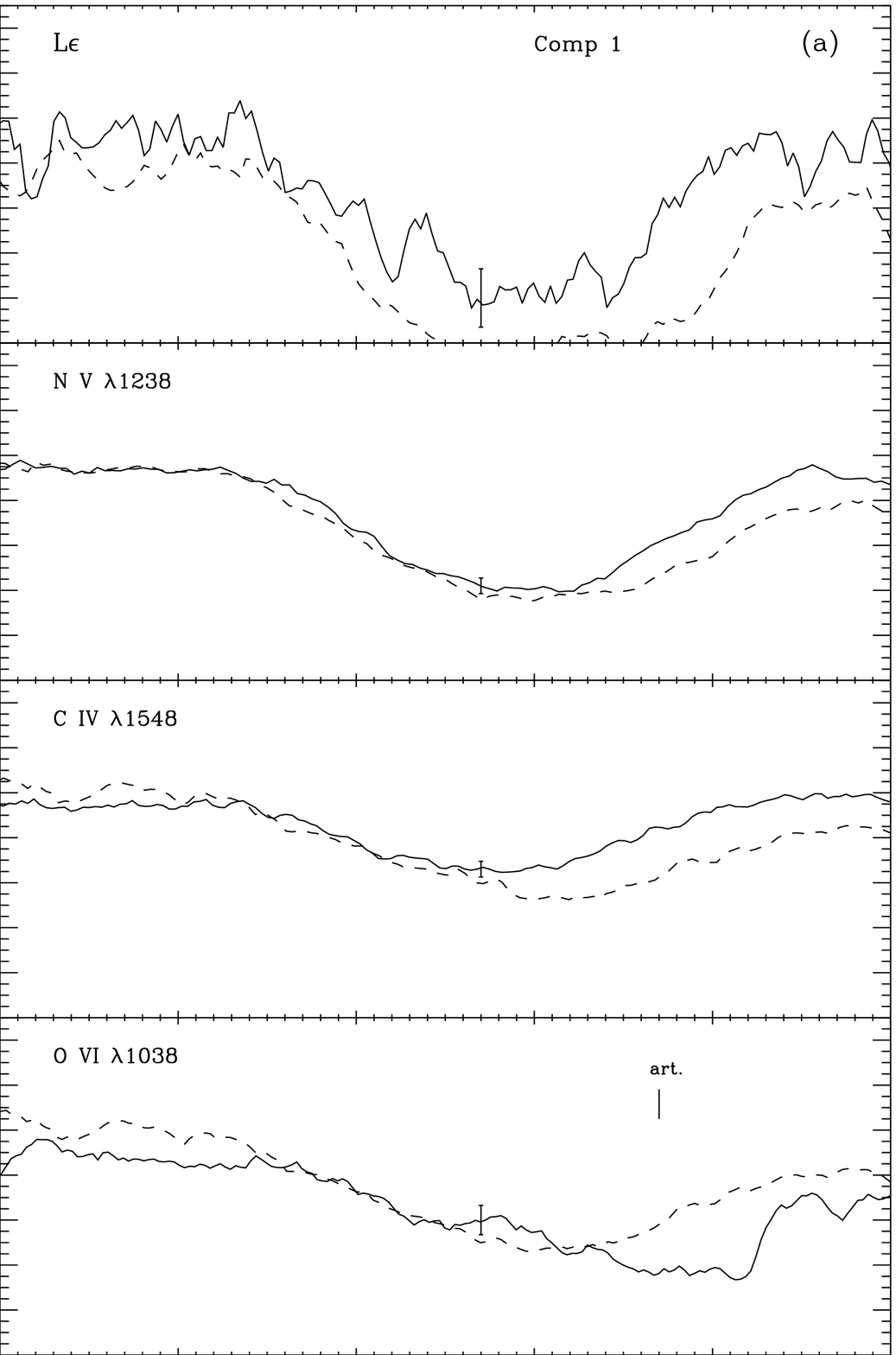}
\caption{Normalized absorption and unobscured flux profiles.  The normalized
absorption profiles for component 1 (5a) and components 2 and 3 (5b) are plotted as solid lines.  
The unobscured normalized flux levels, 1$-C_f$, were computed using our $C_f^c$ and $C_f^l$ 
profiles derived for \ion{H}{1} and are plotted as dashed lines.  Typical error bars are
marked on each spectrum with a vertical line.  Galactic absorption has 
been divided out of each spectrum.  The apparent absorption in the red wing of \ion{O}{6} 
component 1 at $v_r\sim -$1200 km s$^{-1}$ is a detector artifact. \label{fig5}}
\end{figure}

\addtocounter{figure}{-1}

\clearpage
\begin{figure}
\epsscale{0.7}
\plotone{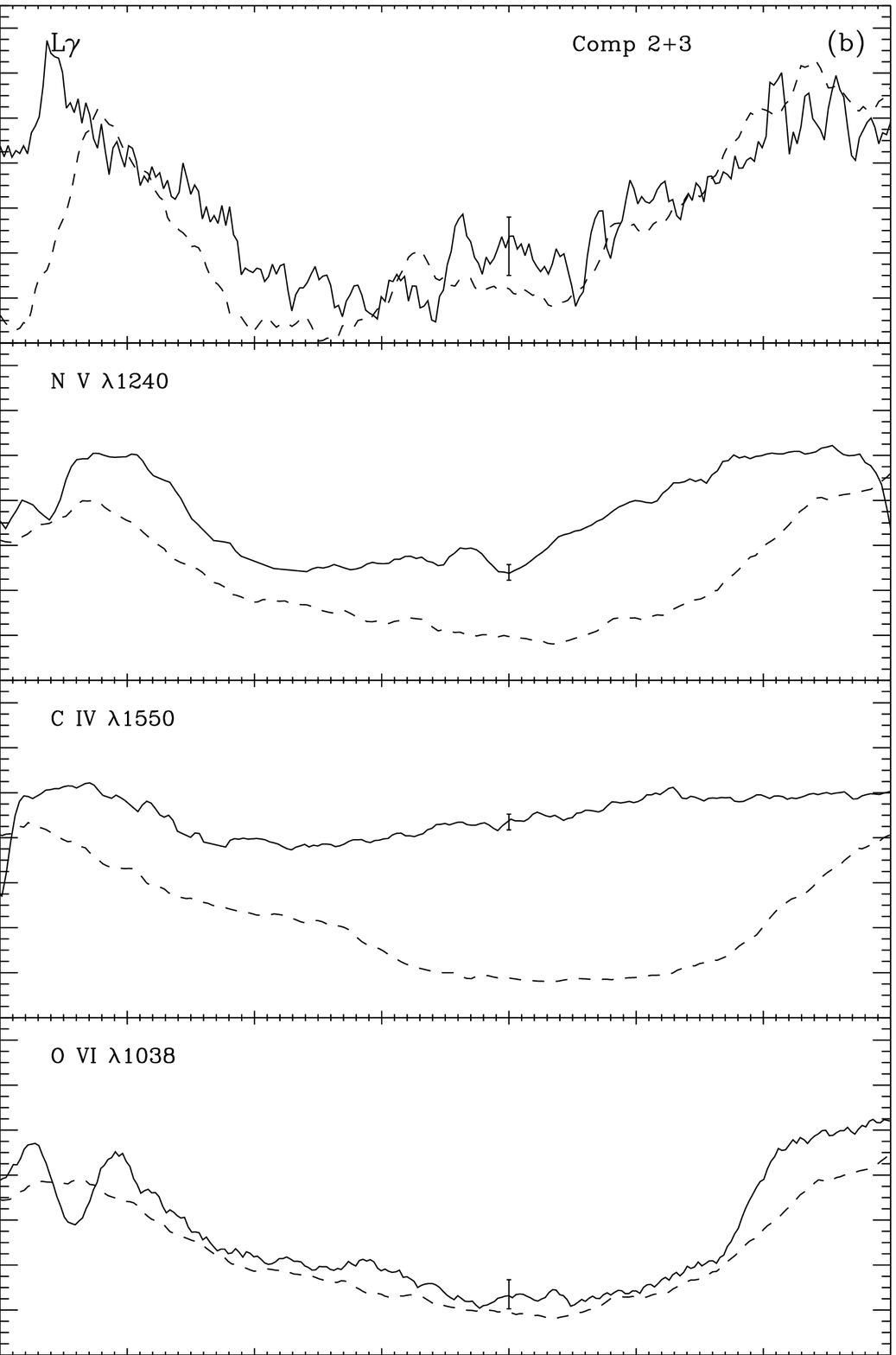}
\caption{Continued \label{fig5b}}
\end{figure}

\end{document}